\documentclass{article}

\begin{document}

\title{Information Integration\\
in Elementary Cellular Automata}

\author{K\'atia~K.~Cassiano\\
Valmir~C.~Barbosa\thanks{Corresponding author (valmir@cos.ufrj.br).}\\
\\
Programa de Engenharia de Sistemas e Computa\c c\~ao, COPPE\\
Universidade Federal do Rio de Janeiro\\
Caixa Postal 68511\\
21941-972 Rio de Janeiro - RJ, Brazil}

\date{}

\maketitle

\begin{abstract}
We study the emergence of information integration in cellular automata (CA) with
respect to states in the long run. Information integration is in this case
quantified by applying the information-theoretic measure known as total
correlation to the long-run distribution of CA states. Total correlation is the
amount by which the total uncertainty associated with cell states surpasses the
uncertainty of the CA state taken as a whole. It is an emergent property, in the
sense that it can only be ascribed to how the cells interact with one another,
and has been linked to the rise of consciousness in the brain. We investigate
total correlation in the evolution of elementary CA for all update rules that
are unique with respect to negation or reflection. For each rule we consider the
usual, deterministic CA behavior, assuming that the initial state is chosen
uniformly at random, and also the probabilistic variant in which every cell, at
all time steps and independently of all others, disobeys the rule's prescription
with a fixed probability. We have found rules that generate as much total
correlation as possible, or nearly so, particularly in Wolfram classes 2 and 3.
We conjecture that some of these rules can be used as CA models of information
integration.

\end{abstract}

\newpage
\section{Introduction}

Given a dynamical system comprising interacting components whose behavior may
lead to uncertain outcomes, a useful tool to quantify such uncertainty is the
well-known information-theoretic notion of entropy. Entropy measures can be
applied to the system in question both at the level of its individual components
and globally, that is, at the level of the system's global (rather than local)
states. Both approaches rely on probability distributions that are interrelated,
as computing the system's global entropy relies on a joint distribution whose
marginals are precisely the distributions to be used to compute the
component-wise entropies. Informally, we say that information is being
integrated by such a system when the sum of all the local entropies surpasses
the global entropy. Because equality between the two quantities can only occur
when all components are independent of one another, such excess entropy implies
that the system as a whole is capable of generating more information than all
the individual components put together.

The emergence of such information surplus cannot be linked directly to the
operation of any one of the system's components, but resides, much less
concretely, in the manner of their interaction. Owing to peculiar
characteristics such as this, information integration has found its way into the
select group of explanatory theories for phenomena such as consciousness. In
fact, one of the leading candidates to explain the rise of consciousness out of
the functioning of the brain is the integrated-information theory \cite{bt08},
based precisely on the generation of excess information such as we have
described. This theory is currently favored by leading neuroscientists
\cite{k12}, in a clear indication that its allure far outweighs its many
theoretical and computational difficulties \cite{nb10}.

Here we study the integration of information in cellular automata (CA) both of
the standard, deterministic variety and of the probabilistic one. In either case
we target the generation of information by the CA in the long run, understood as
information regarding the long-run state of the CA. We assume that the initial
CA state is chosen uniformly at random, which immediately endows the CA's
long-run evolution with some degree of uncertainty even in the deterministic
case. Moreover, our probabilistic CA are characterized by a single parameter
that gives, at all time steps and for each cell independently of the others, the
probability with which the update rule in use is to be disobeyed. Probabilistic
CA are relatively commonplace in CA studies (e.g., to model spin lattices in
statistical physics \cite{dk84,bg85,gjh85,gl89,lms90}), and in this regard, we
note that our model is the same we used previously in an immunity-related study
\cite{cb14}. It is similar to other probabilistic-CA models in the literature
(cf.\ \cite{sc13} and references therein), but notwithstanding this our emphasis
on long-run state probabilities rather than on spatiotemporal patterns sets
apart the specific use of make of it.

Given a CA update rule, our investigation starts with computing the long-run
probability that the CA is found at each possible state. This can be a simple
task in the deterministic case or for a very small number of cells in the
probabilistic case, but can also be painstakingly time-consuming for
probabilistic CA only slightly larger than ten cells. Once the desired
probabilities are known, we compute the information-theoretic measure known as
total correlation, which is defined precisely as the entropy difference alluded
to above and therefore tackles the issue of information integration directly.
We focus our study on elementary CA. Although these are the simplest CA
imaginable, already for them we find rules that promote significant levels of
information integration. This, we believe, bears further witness to the
remarkable capability these simple models can have to capture the essentials of
so many relevant, complex phenomena.

The following is how we proceed. First we present our CA, with particular
emphasis on how to compute their long-run state probabilities, in
Section~\ref{model}. Then, in Section~\ref{ii}, we present all the necessary
information-theoretic notions, including those of entropy, information gain, and
total correlation. Some of the computational difficulties associated with
calculating long-run state probabilities for the probabilistic CA are discussed
briefly in Section~\ref{markov}, after which we present our results in
Section~\ref{results}. Discussion follows in Section~\ref{disc}, then conclusion
in Section~\ref{concl}. We note, before proceeding, that all the theory given
in Sections~\ref{model} and~\ref{ii}, though presented for binary CA, can be
extended to non-binary CA in a straightforward manner. We refrain from the
more general presentation for the sake of notational simplicity only.

\section{Model}\label{model}

We consider binary CA, that is, CA in which each cell's state is either $0$ or
$1$. If each cell has $\delta$ neighbors, then a cell's state at the next
instant of discrete time is a function of its own current state and of the
states of its $\delta$ neighbors. This function is the CA update rule, which in
general can be thought of as one of the $2^{2^{\delta+1}}$ possible tables
having $2^{\delta+1}$ entries, each entry indexed by $\delta+1$ bits.

A binary CA such as this has $2^n$ possible states, where $n$ is the number of
cells, each state corresponding to a member of the set $\{0,1\}^n$. Starting at
some initial state $i_0\in\{0,1\}^n$, at all times thereafter the states of all
cells evolve in lockstep based on the same update rule. What ensues is a
deterministic evolution of the CA state inside the set $\{0,1\}^n$, which by
finiteness must eventually become periodic. This implies that, given the update
rule, the state set $\{0,1\}^n$ can be partitioned into attractor basins, i.e.,
subsets of states to which the evolution of the CA is perpetually confined.

For a fixed update rule and a given value of $n$, we let $\mathcal{B}$ denote
the corresponding set of attractor basins that partitions $\{0,1\}^n$. For each
$B\in\mathcal{B}$, we let $A_B\subseteq B$ be the attractor itself, that is, the
set of periodic states lying at the core of $B$.

Our study in this paper is based on the probability $\sigma_i$ that, having
started at some state chosen uniformly at random, the CA is found in state
$i$ in the long run, that is, as time grows without bounds. Clearly, given the
deterministic evolution scenario outlined thus far this probability depends on
state $i$ being in attractor $A_B$ for some $B\in\mathcal{B}$. That is, we have
\begin{equation}\label{eq:det}
\sigma_i=\left\{
\begin{array}{ll}
\sigma_B/\vert A_B\vert,
&\mbox{if $i\in A_B$ for some $B\in\mathcal{B}$;}\\
0,&\mbox{otherwise},
\end{array}
\right.
\end{equation}
where $\sigma_B$ is the size of basin $B$ relative to the entire set
$\{0,1\}^n$, i.e., $\sigma_B=\vert B\vert/2^n$.

In order to see that Equation~(\ref{eq:det}) does indeed hold, it suffices that
we write $\sigma_i$ as the total probability
\begin{equation}
\sigma_i=\sum_{i_0\in\{0,1\}^n}\sigma_{i\vert i_0}\mathrm{Pr}(i_0),
\end{equation}
where $\sigma_{i\vert i_0}$ is the conditional probability that the CA is found
at state $i$ in the long run, given that its initial state was $i_0$, and
$\mathrm{Pr}(i_0)$ is the probability that it did start at $i_0$. We then simply
recognize that
\begin{equation}
\sigma_{i\vert i_0}=\left\{
\begin{array}{ll}
1/\vert A_B\vert,
&\mbox{if $i\in A_B$ for $B\in\mathcal{B}$ such that $i_0\in B$;}\\
0,&\mbox{otherwise}
\end{array}
\right.
\end{equation}
and that $\mathrm{Pr}(i_0)=1/2^n$.

In this standard, deterministic scenario, whatever uncertainty there is
regarding the long-run state of the CA is a consequence of the method to
determine the initial state $i_0$ (uniformly at random). In the second scenario
that we explore, we replace this source of uncertainty by adding randomness to
the CA dynamics itself. We do this by letting each cell, independently of all
others and at every time step, behave differently than the update rule in use
mandates with a given probability. We describe the resulting probabilistic CA
next.

Let $x$ denote a cell's next state and let $b\in\{0,1\}$ be the update rule's
current prescription for the value of $x$. Given a probability parameter $p$,
what we do is to let
\begin{equation}
x:=\left\{
\begin{array}{ll}
1-b,&\mbox{with probability $p$;}\\
b,&\mbox{otherwise.}
\end{array}
\right.
\end{equation}
Clearly, the deterministic scenario is recovered by letting $p=0$. For $p>0$,
what is now happening in CA-wide terms is that any CA state can be reached from
any other in a single step, regardless of being members of the same basin or, if
they are members of the same basin, of how they are positioned inside it with
respect to each other. In particular, a basin's attractor is no longer
inescapable.

If $k_i$ denotes the deterministic successor of state $i$ in the CA dynamics,
then the one-step transition probability from state $i$ to state $j$, call it
$p_{i,j}$, is given by
\begin{equation}\label{eq:tprob1}
p_{i,j}=p^{D_{j,k_i}}(1-p)^{n-D_{j,k_i}},
\end{equation}
where $D_{j,k_i}$ is the Hamming distance between states $j$ and $k_i$, that is,
the number of cells at which they differ. What this expression is saying is
that, in order for $j$ to be obtained from $i$ in a single time step, it is
necessary and sufficient that both all cells at which $j$ and $k_i$ differ
disobey the update rule (independently of one another, each with probability
$p$) and none of the others do (independently of one another, each with
probability $1-p$). We can rewrite $p_{i,j}$ as
\begin{equation}\label{eq:tprob2}
p_{i,j}=(1-p)^n\left(\frac{1-p}{p}\right)^{-D_{j,k_i}},
\end{equation}
which for $0<p<0.5$ emphasizes its exponential decay with $D_{j,k_i}$ and
highlights the fact that the one-step transition from $i$ to $k_i$, the same
that takes place in the deterministic case, is still the most probable one,
being in fact exponentially more likely than any other. The probability of this
transition is obtained by letting $j=k_i$, whence $D_{j,k_i}=0$ and
$p_{i,k_i}=(1-p)^n$.

In the probabilistic case, we denote by $\pi_i$ the probability that, having
started at some state chosen uniformly at random, the CA is found in state $i$
in the long run. As before, we can express $\pi_i$ as the total probability
\begin{equation}\label{eq:taut}
\pi_i=\sum_{i_0\in\{0,1\}^n}\pi_{i\vert i_0}\mathrm{Pr}(i_0),
\end{equation}
where $\pi_{i\vert i_0}$ is the long-run conditional probability that the CA is
found at state $i$, given that it started at state $i_0$. However, it can be
easily argued that, provided $p>0$, the CA's long-run behavior is independent
of $i_0$, so $\pi_{i\vert i_0}=\pi_i$ and Equation~(\ref{eq:taut}) turns out to
be no more than a tautology ($\pi_i=\pi_i$).

In fact, for $p>0$ the CA dynamics gives rise to a discrete-time Markov chain of
transition-probability matrix $P=[p_{i,j}]$. This can be easily verified by
simply checking that, in $P$, all the elements in any row add up to $1$ (i.e.,
$P$ is stochastic):
\begin{equation}\label{eq:stoch}
\sum_{j\in\{0,1\}^n}p_{i,j}=\sum_{d=0}^n{n\choose d}p^d(1-p)^{n-d}=1.
\end{equation}
Because $p_{i,j}>0$ regardless of $i$ or $j$, this chain is ergodic and,
moreover, the stationary probability that it entails for state $i$ is precisely
$\pi_i$, no matter what the initial state may have been. Our long-run
probabilities can then be found by solving the system $\pi=\pi P$, where
$\pi=[\pi_i]$ is a row vector.

We finalize the section with the introduction of marginal versions of
probabilities $\sigma_i$ and $\pi_i$. They are marginal in the sense of relating
to one cell exclusively (as opposed to $\sigma_i$ or $\pi_i$, which relate to
all cells concomitantly). For $c=1,2,\ldots,n$, we denote by $\sigma_{c,b}$ the
long-run probability that, in the deterministic case, the state of cell $c$ is
$b\in\{0,1\}$. Clearly,
\begin{equation}\label{eq:dmprob}
\sigma_{c,b}=
\sum_{i\in S_{c,b}}\sigma_i=
\sum_{B\in\mathcal{B}}
\frac{\sigma_B\vert S_{c,b}\cap A_B\vert}{\vert A_B\vert},
\end{equation}
where $S_{c,b}\subset\{0,1\}^n$ is the set of all CA states in which cell $c$
has state $b$. Likewise, the corresponding probability in the probabilistic case
is denoted by $\pi_{c,b}$ and given by
\begin{equation}
\pi_{c,b}=\sum_{i\in S_{c,b}}\pi_i.
\end{equation}

\section{Information integration}\label{ii}

In both the deterministic and the probabilistic scenarios discussed in
Section~\ref{model}, uncertainty regarding the long-run state of the CA has a
role to play. In the deterministic scenario, this uncertainty owes to the fact
that the initial CA state $i_0$ may be any state. In the probabilistic scenario,
by contrast, it stems from the nondeterministic mechanics that now underlies the
CA's workings. Moreover, because of the system's underlying Markovian nature, in
this scenario fixing $i_0$ has no effect on long-run uncertainty, which depends
exclusively on the probability parameter $p$, if nonzero.

In this section we discuss the information-theoretic tools that will be needed
to characterize such uncertainty, its relation to how much uncertainty there can
be at most, and also its role in highlighting how independent the various cells
are of one another given an update rule and the value of $n$. We discuss these
tools in terms of generic CA-state probabilities $\rho_i$ (which stand for
either $\sigma_i$ or $\pi_i$, as the case may be) or likewise generic
cell-state probabilities $\rho_{c,b}$ (placeholders for either $\sigma_{c,b}$ or
$\pi_{c,b}$).

Our main tool is the well-known Shannon entropy, which measures the uncertainty
of a set of random variables, given a joint probability distribution of their
values. The random variables of interest to us, call them $X_1,X_2,\ldots,X_n$,
are those describing the various cells' states. The corresponding joint
distribution is any set of probabilities over the set $\{0,1\}^n$, provided they
add up to $1$. Denoting this joint entropy by $H$, we have
\begin{equation}\label{eq:jentr}
H=-\sum_{i\in\{0,1\}^n}
\mathrm{Pr}(\mbox{CA state is $i$})
\log_2\mathrm{Pr}(\mbox{CA state is $i$}),
\end{equation}
where the logarithm to the base $2$ is meant to let $H$ be expressed in
information-theoretic bits.

The value of $H$ is maximized by the distribution that expresses the greatest
possible uncertainty, that is, the distribution for which
$\mathrm{Pr}(\mbox{CA state is $i$})=1/2^n$ for every CA state $i$. The
resulting maximum value of $H$ is therefore $n$, which works as the absolute
upper bound against which we compare long-run uncertainties in both the
deterministic and probabilistic cases. The resulting difference, which we refer
to as information gain, is here denoted by $G^\rho$ and given by
\begin{equation}\label{eq:jgain}
G^\rho=n-H^\rho,
\end{equation}
where $H^\rho$ results from Equation~(\ref{eq:jentr}) by substituting $\rho_i$
for $\mathrm{Pr}(\mbox{CA state is $i$})$:
\begin{equation}
H^\rho=-\sum_{i\in\{0,1\}^n}\rho_i\log_2\rho_i.
\end{equation}

A marginal version of the entropy $H$ can be defined for each cell $c$ to
quantify the uncertainty associated with its state. This new entropy is denoted
by $H_c$ and given by
\begin{equation}\label{eq:mentr}
H_c=-\sum_{b\in\{0,1\}}
\mathrm{Pr}(\mbox{state of cell $c$ is $b$})
\log_2\mathrm{Pr}(\mbox{state of cell $c$ is $b$}).
\end{equation}
The value of $H_c$ is maximized when
$\mathrm{Pr}(\mbox{state of cell $c$ is $b$})=0.5$ regardless of $b$, which
leads to a maximum value of $1$. As in the case of $H$, this is the absolute
upper bound against which we compare the long-run uncertainty associated with
the state of cell $c$. The resulting information gain, which we denote by
$G^\rho_c$, is given by
\begin{equation}\label{eq:mgain}
G^\rho_c=1-H^\rho_c,
\end{equation}
where $H^\rho_c$ is obtained from Equation~(\ref{eq:mentr}) by replacing
$\mathrm{Pr}(\mbox{state of cell $c$ is $b$})$ by $\rho_{c,b}$:
\begin{equation}
H^\rho_c=-\sum_{b\in\{0,1\}}\rho_{c,b}\log_2\rho_{c,b}.
\end{equation}

If the random variables $X_1,X_2,\ldots,X_n$ are all independent of one another
given the $\rho_i$'s, i.e., if $\rho_i=\prod_{c=1}^n\rho_{c,b_c}$ for all
$i\in\{0,1\}^n$, where $b_c$ is the state of cell $c$ in CA state $i$, then it
follows from the expressions for $H^\rho$ and $H^\rho_c$ that
$\sum_{c=1}^nH^\rho_c=H^\rho$. Only for independent random variables does this
happen. In all other cases we have $\sum_{c=1}^nH^\rho_c>H^\rho$, where the two
sides differ by what is known as the total correlation among the $n$ variables
\cite{w60}. We denote total correlation by $C^\rho$, which is then given by
\begin{equation}\label{eq:tcorr}
C^\rho=\sum_{c=1}^nH^\rho_c-H^\rho.
\end{equation}
(For $n=2$, total correlation is also known as the mutual information between
the two variables \cite{h80}.)

An interesting interpretation of total correlation comes from rewriting
Equation~(\ref{eq:tcorr}) in terms of the information gains $G^\rho$ (for the
entire CA) and $G^\rho_c$ (for cell $c$). By Equations~(\ref{eq:jgain})
and~(\ref{eq:mgain}), we have $C^\rho=G^\rho-\sum_{c=1}^nG^\rho_c$, whence
\begin{equation}\label{eq:fgain}
G^\rho=\sum_{c=1}^nG^\rho_c+C^\rho.
\end{equation}
That is, total correlation is the amount of information gain that the CA's
evolution in time produces in excess of the total information gain that is
already produced at the level of the cells. We refer to the fraction of
information gain that corresponds to total correlation as a total-correlation
ratio, denoted by $r^\rho$:
\begin{equation}\label{eq:fratio}
r^\rho=\frac{C^\rho}{G^\rho}.
\end{equation}
In our analyses, $G^\rho$ and $r^\rho$ are used as the premier indicators of
information integration.

\section{Computational issues}\label{markov}

For each update rule of interest, and for fixed values of $n$ and $p$,
computing $G^\sigma$, $r^\sigma$, $G^\pi$, and $r^\pi$ requires that the
long-run probabilities $\sigma_i$ and $\pi_i$ be found for every CA state $i$.
There are two computational challenges related to obtaining these probabilities.
The first one affects both the deterministic case (to which the $\sigma_i$'s
refer) and the probabilistic case (to which the $\pi_i$'s refer), and has to do
with mapping out all $2^n$ CA states onto the basin-of-attraction field. This
can be challenging because, depending on the value of $n$, substantial amounts
of main storage may be needed.

The second, and more serious, computational challenge has to do with the time
required to find the $\pi_i$'s. As we discussed in Section~\ref{model}, this
amounts to solving the system $\pi=\pi P$, subject to the constraints that
$\pi_i>0$ for all $i\in\{0,1\}^n$ and $\sum_{i\in\{0,1\}^n}\pi_i=1$. Since $P$
is a $2^n\times 2^n$ matrix of strictly positive elements and possessing no
known symmetries or some other type of structure that might simplify
calculations, solving this system tends to be quite burdensome even for modest
values of $n$. We have used state-of-the-art solution techniques through the
solver that is freely available as part of the Tangram-II modeling tool
\cite{slsrdfjm06}, but even so only for $n<13$ has the solution of the system
proven feasible. Based on preliminary experiments with $n=13$, we estimate that
solving a single instance of the system in this case would require about two
months on an Intel Xeon E5-1650 running at 3.2GHz with enough main storage to
preclude the need for any accesses to secondary storage.

\section{Results}\label{results}

We give results for the so-called elementary CA, that is, one-dimensional CA
with neighborhood size $\delta=2$, and adopt periodic boundaries in all cases
(i.e., the first and last cells in all CA are neighbors of each other).
Elementary CA admit $256$ distinct update rules and here we use the standard
Wolfram numbering system \cite{w83} in referring to them.

Of these $256$ rules, only $88$ are unique in the sense of how the resulting
basins are structured. Given any one of these $88$ rules, say $R$, any other
rule $R'$ satisfying the property we give next can be identified amid the
remaining $168$ rules. The property in question is that a mapping $g$ between CA
states exists such that $R$ leads from CA state $i$ to its deterministic
successor $k_i$ if and only if $R'$ leads from $g(i)$ to $g(k_i)$. The two
mappings that we use are negation [adding a cell's state in $i$ to its state in
$g(i)$ yields $1$] and reflection [the state of cell $c$ in $i$ is the same as
the state of cell $n-c+1$ in $g(i)$]. Our $88$ unique rules are such that no two
of them are equivalent to each other by negation or reflection. This criterion
alone leads to several satisfying sets of $88$ rules. Our choice has been to
follow Wuensche and Lesser \cite{wl92}, who in their atlas group all rules into
equivalence classes of at most 4 rules or into larger clusters of at most 8
rules as equivalence classes of pairwise complementary rules are joined. We
select for inclusion in the group of 88 the least-number rule of each larger
cluster, along with its complement if not already in the first rule's
equivalence class.

The identical structuring of basins for two rules that are equivalent by
negation or reflection provides sufficient justification for eliminating one of
them when handling the deterministic case of Section~\ref{model}. In the
probabilistic case, eliminating one of the two rules from consideration on the
basis of the equivalence of results requires, in addition, that the transition
probabilities $p_{i,j}$ and $p_{g(i),g(j)}$ be the same for any two CA states
$i$ and $j$. To see that this does indeed hold, notice that it follows directly
from Equation~(\ref{eq:tprob1}), since $D_{j,k_i}=D_{g(j),g(k_i)}$ when $g$
stands for negation or reflection.

Another curious property stemming from the definition of $p_{i,j}$ in
Equation~(\ref{eq:tprob1}) is the following. Recall that two rules are
complementary to each other when, given the same input, one of them outputs bit
$b$ if and only if the other outputs $1-b$. So, for example, if letting $p=0$ in
the probabilistic case reproduces the deterministic behavior of the underlying
update rule, say $R$, then letting $p=1$ for the same underlying rule $R$ also
induces deterministic behavior, but of the rule $R'$ that is complementary to
$R$. Something similar occurs when $p>0$. Given any CA state $i$ and the
underlying rule $R$ that determines $i$'s deterministic successor, $k_i$, the
probability that $i$ is followed by $j$ when $R$ is disobeyed at each cell
independently with probability $p$ is the $p_{i,j}$ of
Equation~(\ref{eq:tprob1}). Should we use $R'$ instead and let it be disobeyed
at each cell independently with probability $1-p$, the transition probability
from $i$ to $j$ would be $(1-p)^{D_{j,k'_i}}p^{n-D_{j,k'_i}}$, where $k'_i$ is
the deterministic successor of CA state $i$ under $R'$. But it so happens that
$D_{j,k_i}+D_{j,k'_i}=n$, so this probability can be rewritten as
$(1-p)^{n-D_{j,k_i}}p^{D_{j,k_i}}$, which is none other than the very same
$p_{i,j}$ with which the transition from $i$ to $j$ occurs given $R$ and $p$.
This means that it makes no sense to seek results for both $p<0.5$ and $p>0.5$.
After all, working with $p>0.5$ for some rule $R$ in the group of $88$ is the
same as doing it with probability $1-p<0.5$ for the rule $R'$ that is
complementary to $R$. Rule $R'$, in turn, either is one of the $88$ itself (and
is then already covered) or is not (in which case it is equivalent to some rule
in the group of $88$ and, again, is already covered). We then use $p<0.5$
exclusively.

Our results are summarized in Tables~\ref{n=11} and~\ref{n=12}, respectively
for $n=11$ and $n=12$ cells. Each table contains information gains and
total-correlation ratios for all $88$ rules, each rule identified as noted above
alongside its Wolfram class (1 through 4) \cite{w84}. All data are given for the
deterministic case (identified as the $p=0$ case) and two probabilistic cases,
viz.\ with probabilities $p=0.001$ and $p=0.01$. In either table, $G^\sigma$ and
$G^\pi$ are obtained by substituting distribution $\sigma$ or $\pi$,
respectively, for the $\rho$ on which information gain is defined [cf., e.g.,
Equation~(\ref{eq:fgain})]. The same holds for $r^\sigma$ and $r^\pi$ with
respect to the total-correlation ratio given in Equation~(\ref{eq:fratio}).

\begin{table}[p]
\caption{Information gains and total-correlation ratios for $n=11$.}
\centering
\begin{tabular}{|c|c|cc|cc|cc|}
\hline
& &\multicolumn{2}{|c|}{$p=0$} &\multicolumn{2}{|c|}{$p=0.001$} &\multicolumn{2}{|c|}{$p=0.01$}\\
\cline{3-8}
Rule &W.\ cl.\footnotemark[1] &$G^\sigma$ &$r^\sigma$ &$G^\pi$ &$r^\pi$ &$G^\pi$ &$r^\pi$\\
\hline
\it   0  &\it 1 &\it 11.0000  &\it 0.0000 &\it 10.8745  &\it 0.0000 &\it 10.1113  &\it 0.0000\\
\it 248  &\it 1  &9.8149  &0.0972 &\it 10.8743  &\it 0.0000 &\it 10.0962  &\it 0.0000\\
\it 249  &\it 1 &10.8664  &0.0439 &\it 10.8743  &\it 0.0000 &\it 10.0964  &\it 0.0000\\
\it 250  &\it 1 &10.9939  &0.0055 &\it 10.8744  &\it 0.0000 &\it 10.1039  &\it 0.0000\\
\it 251  &\it 1 &\it 11.0000  &\it 0.0000 &\it 10.8744  &\it 0.0000 &\it 10.1040  &\it 0.0000\\
\it 252  &\it 1 &10.9939  &0.0055 &\it 10.8744  &\it 0.0000 &\it 10.1039  &\it 0.0000\\
\it 253  &\it 1 &\it 11.0000  &\it 0.0000 &\it 10.8744  &\it 0.0000 &\it 10.1041  &\it 0.0000\\
\it 254  &\it 1 &10.9939  &0.0055 &\it 10.8745  &\it 0.0000 &\it 10.1112  &\it 0.0000\\
\hline
  1  &2  &5.4737  &0.9662  &3.9532  &0.9269  &3.5371  &0.9132\\
  2  &2  &6.9611  &0.0888  &5.0112  &0.2350  &4.6206  &0.2016\\
  3  &2  &3.1524  &0.9650  &2.5157  &0.9455  &2.2041  &0.9344\\
  4  &2  &5.4158  &0.1435  &4.2273  &0.1342  &3.9721  &0.1121\\
  5  &2  &2.8336  &0.9258  &2.5157  &0.9455  &2.2041  &0.9344\\
  6  &2  &6.6380  &0.1423  &4.2454  &0.4671  &3.5261  &0.4197\\
  7  &2  &5.5036  &0.9945  &9.7352  &1.0000  &8.0350  &0.9998\\
  9  &2  &9.2894  &0.9998  &5.4167  &0.9504  &3.7860  &0.9174\\
 10  &2  &3.9120  &0.2753  &3.3973  &0.3915  &3.1046  &0.3591\\
\bf  11  &\bf 2  &\bf 4.4633  &\bf 1.0000  &4.3969  &0.9978  &3.4942  &0.9964\\
 12  &2  &3.5437  &0.3919  &3.3973  &0.3915  &3.1046  &0.3591\\
 13  &2  &6.7514  &0.9684  &7.3462  &0.9910  &6.2310  &0.9890\\
 14  &2  &6.4286  &0.4749  &5.5259  &0.9930  &4.3577  &0.9873\\
\bf  15  &\bf 2  &\bf 0.0338  &\bf 1.0000  &\bf 0.0000  &\bf 1.0000  &\bf 0.0000  &\bf 1.0000\\
\hline
\end{tabular}

{\footnotesize\footnotemark[1]Wolfram class.\hfill}
\label{n=11}
\end{table}

\addtocounter{table}{-1}
\begin{table}[p]
\centering
\caption{Continued.}
\begin{tabular}{|c|c|cc|cc|cc|}
\hline
& &\multicolumn{2}{|c|}{$p=0$} &\multicolumn{2}{|c|}{$p=0.001$} &\multicolumn{2}{|c|}{$p=0.01$}\\
\cline{3-8}
Rule &W.\ cl.\footnotemark[1] &$G^\sigma$ &$r^\sigma$ &$G^\pi$ &$r^\pi$ &$G^\pi$ &$r^\pi$\\
\hline
\bf  19  &\bf 2  &\bf 4.0612  &\bf 1.0000  &9.3448  &1.0000  &6.4186  &0.9999\\
\bf  23  &\bf 2  &\bf 4.4155  &\bf 1.0000  &\bf 9.6667  &\bf 1.0000  &\bf 7.7807  &\bf 1.0000\\
 24  &2  &5.9045  &0.1561  &5.0244  &0.2166  &4.4889  &0.1796\\
 25  &2  &8.7170  &0.9998  &5.1766  &0.9869  &3.2991  &0.9732\\
 26  &2  &3.2929  &0.5160  &1.8577  &0.7354  &1.4591  &0.6972\\
 27  &2  &2.5655  &0.9892  &2.0429  &0.9860  &1.6168  &0.9835\\
 28  &2  &5.5200  &0.9661  &6.2799  &1.0000  &4.8558  &0.9998\\
\bf  29  &\bf 2  &1.5038  &0.9926  &\bf 1.5355  &\bf 1.0000  &\bf 1.3491  &\bf 1.0000\\
 33  &2  &3.9746  &0.8913  &3.6030  &0.8923  &3.0633  &0.8663\\
 35  &2  &3.5095  &0.9706  &2.8857  &0.8440  &2.3023  &0.8222\\
 36  &2  &7.4540  &0.0688  &6.0282  &0.0638  &5.5462  &0.0456\\
 37  &2  &4.9850  &0.9517  &4.1290  &0.9427  &2.6269  &0.9331\\
 38  &2  &3.6690  &0.3740  &2.8163  &0.4666  &2.4383  &0.4259\\
 41  &2  &4.4215  &0.9244  &3.3402  &0.8589  &2.1464  &0.7690\\
\bf  43  &\bf 2  &3.9427  &0.9999  &\bf 5.0562  &\bf 1.0000  &\bf 4.0296  &\bf 1.0000\\
 46  &2  &5.9045  &0.6557  &4.9939  &0.9625  &4.2328  &0.9580\\
 50  &2  &4.4377  &0.9948  &6.3393  &1.0000  &5.1962  &0.9999\\
\bf  51  &\bf 2  &\bf 0.0000  &\bf 1.0000  &\bf 0.0000  &\bf 1.0000  &\bf 0.0000  &\bf 1.0000\\
\bf  57  &\bf 2  &7.5139  &0.9977  &\bf 6.2250  &\bf 1.0000  &\bf 4.6150  &\bf 1.0000\\
 58  &2  &8.1430  &0.6769  &7.1978  &0.9193  &5.4153  &0.9156\\
 62  &2  &3.8652  &0.9960  &3.0641  &0.8693  &1.9817  &0.8497\\
\bf  77  &\bf 2  &4.3581  &0.9814  &\bf 6.4074  &\bf 1.0000  &\bf 5.6031  &\bf 1.0000\\
\hline
\end{tabular}

{\footnotesize\footnotemark[1]Wolfram class.\hfill}
\end{table}

\addtocounter{table}{-1}
\begin{table}[p]
\centering
\caption{Continued.}
\begin{tabular}{|c|c|cc|cc|cc|}
\hline
& &\multicolumn{2}{|c|}{$p=0$} &\multicolumn{2}{|c|}{$p=0.001$} &\multicolumn{2}{|c|}{$p=0.01$}\\
\cline{3-8}
Rule &W.\ cl.\footnotemark[1] &$G^\sigma$ &$r^\sigma$ &$G^\pi$ &$r^\pi$ &$G^\pi$ &$r^\pi$\\
\hline
 94  &2  &4.2574  &0.9567  &4.5916  &0.9291  &3.5605  &0.9225\\
\bf 178  &\bf 2  &4.4372  &0.9949  &\bf 6.4074  &\bf 1.0000  &\bf 5.6031  &\bf 1.0000\\
197  &2  &6.8246  &0.9860  &7.3036  &0.9909  &5.9481  &0.9880\\
\bf 198  &\bf 2  &5.5163  &0.9215  &\bf 6.3246  &\bf 1.0000  &\bf 5.0879  &\bf 1.0000\\
201  &2  &2.4247  &0.5599  &2.4971  &0.4695  &2.0955  &0.4368\\
\bf 204  &\bf 2  &\bf 0.0000  &\bf 1.0000  &\bf 0.0000  &\bf 1.0000  &\bf 0.0000  &\bf 1.0000\\
205  &2  &1.6579  &0.5761  &1.6152  &0.6299  &1.4590  &0.6054\\
\bf 210  &\bf 2  &0.2646  &0.9252  &\bf 0.0000  &\bf 1.0000  &\bf 0.0000  &\bf 1.0000\\
\bf 212  &\bf 2  &5.6692  &0.5589  &\bf 5.0562  &\bf 1.0000  &\bf 4.0296  &\bf 1.0000\\
214  &2  &8.1190  &0.7069  &4.3917  &0.5350  &3.4123  &0.4811\\
217  &2  &5.2172  &0.2670  &4.9732  &0.2331  &4.3655  &0.1972\\
218  &2  &6.7121  &0.9262  &5.5600  &0.1032  &5.0568  &0.0752\\
220  &2  &3.5427  &0.4380  &3.3903  &0.3906  &3.0595  &0.3541\\
222  &2  &4.3997  &0.6183  &4.0985  &0.1503  &3.8358  &0.1271\\
\bf 226  &\bf 2  &4.3731  &0.3908  &\bf 3.5097  &\bf 1.0000  &\bf 2.9098  &\bf 1.0000\\
227  &2  &4.8292  &0.3590  &3.8792  &0.8679  &3.2172  &0.8445\\
228  &2  &5.3784  &0.2321  &4.9405  &0.2298  &4.1640  &0.1823\\
229  &2  &5.2951  &0.2537  &3.7003  &0.5585  &2.9112  &0.5054\\
230  &2  &6.5927  &0.9993  &5.0247  &0.2164  &4.4858  &0.1779\\
\bf 232  &\bf 2  &4.4155  &0.7805  &\bf 9.6667  &\bf 1.0000  &\bf 7.7807  &\bf 1.0000\\
233  &2  &8.7447  &0.2280 &10.8486  &0.0009  &9.8961  &0.0048\\
236  &2  &3.3083  &0.4426 &10.7658  &0.0082  &9.3477  &0.0448\\
\hline
\end{tabular}

{\footnotesize\footnotemark[1]Wolfram class.\hfill}
\end{table}

\addtocounter{table}{-1}
\begin{table}[p]
\centering
\caption{Continued.}
\begin{tabular}{|c|c|cc|cc|cc|}
\hline
& &\multicolumn{2}{|c|}{$p=0$} &\multicolumn{2}{|c|}{$p=0.001$} &\multicolumn{2}{|c|}{$p=0.01$}\\
\cline{3-8}
Rule &W.\ cl.\footnotemark[1] &$G^\sigma$ &$r^\sigma$ &$G^\pi$ &$r^\pi$ &$G^\pi$ &$r^\pi$\\
\hline
237  &2  &7.4470  &0.4177 &10.8377  &0.0015  &9.8405  &0.0077\\
\bf 240  &\bf 2  &0.0169  &0.9630  &\bf 0.0000  &\bf 1.0000  &\bf 0.0000  &\bf 1.0000\\
241  &2  &1.9915  &0.6295  &1.4947  &0.5414  &1.3441  &0.5100\\
242  &2  &4.2560  &0.9690  &3.3110  &0.5057  &3.0030  &0.4793\\
243  &2  &3.9120  &0.2647  &3.3973  &0.3915  &3.1046  &0.3591\\
244  &2  &2.5174  &0.9963  &2.5485  &0.4707  &2.2923  &0.4393\\
246  &2  &6.1468  &0.9712  &5.0110  &0.2351  &4.6170  &0.2018\\
\hline
 18  &3  &5.1029  &0.6673  &4.5431  &0.6601  &3.4291  &0.5699\\
 22  &3  &4.7351  &0.8053  &4.0192  &0.8740  &2.6205  &0.8076\\
 30  &3  &8.2934  &0.3914  &1.8822  &\underline{1.0000}  &0.4181  &\underline{1.0000}\\
\bf  45  &\bf 3  &4.1107  &0.9999  &\bf 0.0000  &\bf 1.0000  &\bf 0.0000  &\bf 1.0000\\
\bf  60  &\bf 3  &3.7936  &0.7492  &\bf 0.9134  &\bf 1.0000  &\bf 0.5322  &\bf 1.0000\\
 73  &3  &2.5267  &0.9600  &2.4561  &0.9970  &1.8017  &0.9894\\
\bf  90  &\bf 3  &1.1451  &0.9937  &\bf 0.9134  &\bf 1.0000  &\bf 0.5322  &\bf 1.0000\\
\bf 105  &\bf 3  &\bf 0.1451  &\bf 1.0000  &\bf 0.0000  &\bf 1.0000  &\bf 0.0000  &\bf 1.0000\\
126  &3  &5.2229  &0.9925  &4.4457  &0.9797  &3.3005  &0.9715\\
\bf 150  &\bf 3  &0.0726  &0.9715  &\bf 0.0000  &\bf 1.0000  &\bf 0.0000  &\bf 1.0000\\
161  &3  &5.0171  &0.9921  &4.3989  &0.9726  &2.9997  &0.9516\\
182  &3  &5.0748  &0.8158  &4.6099  &0.6379  &3.4097  &0.5480\\
225  &3  &9.7681  &0.0904  &1.7866  &0.9771  &0.3309  &0.9938\\
\hline
 54  &4  &4.7445  &0.9380  &3.5125  &0.9772  &2.1520  &0.9548\\
193  &4  &8.4343  &0.4560  &3.0861  &0.9236  &1.5514  &0.9270\\
\hline
\end{tabular}

{\footnotesize\footnotemark[1]Wolfram class.\hfill}
\end{table}

\begin{table}[p]
\centering
\caption{Information gains and total-correlation ratios for $n=12$.}
\begin{tabular}{|c|c|cc|cc|cc|}
\hline
& &\multicolumn{2}{|c|}{$p=0$} &\multicolumn{2}{|c|}{$p=0.001$} &\multicolumn{2}{|c|}{$p=0.01$}\\
\cline{3-8}
Rule &W.\ cl.\footnotemark[1] &$G^\sigma$ &$r^\sigma$ &$G^\pi$ &$r^\pi$ &$G^\pi$ &$r^\pi$\\
\hline
\it   0   &\it 1 &\it 12.0000  &\it 0.0000 &\it 11.8631  &\it 0.0000 &\it 11.0305  &\it 0.0000\\
\it 248   &\it 1 &10.9633  &0.0772 &\it 11.8629  &\it 0.0000 &\it 11.0140  &\it 0.0000\\
\it 249   &\it 1 &11.9093  &0.0329 &\it 11.8629  &\it 0.0000 &\it 11.0142  &\it 0.0000\\
\it 250   &\it 1 &11.7678  &0.0987 &\it 11.8630  &\it 0.0000 &\it 11.0225  &\it 0.0000\\
\it 251   &\it 1 &11.9934  &0.0027 &\it 11.8630  &\it 0.0000 &\it 11.0225  &\it 0.0000\\
\it 252   &\it 1 &11.9967  &0.0030 &\it 11.8630  &\it 0.0000 &\it 11.0225  &\it 0.0000\\
\it 253   &\it 1 &\it 12.0000  &\it 0.0000 &\it 11.8630  &\it 0.0000 &\it 11.0226  &\it 0.0000\\
\it 254   &\it 1 &11.9967  &0.0030 &\it 11.8631  &\it 0.0000 &\it 11.0304  &\it 0.0000\\
\hline
  1   &2  &6.0118  &0.9647  &4.3124  &0.9269  &3.8585  &0.9132\\
  2   &2  &7.3975  &0.0940  &5.4670  &0.2351  &5.0408  &0.2017\\
  3   &2  &3.4231  &0.9635  &2.7444  &0.9455  &2.4045  &0.9344\\
  4   &2  &5.9082  &0.1469  &4.6116  &0.1342  &4.3332  &0.1121\\
  5   &2  &3.1116  &0.9285  &2.7524  &0.9459  &2.4113  &0.9348\\
  6   &2  &5.1189  &0.3888  &4.5158  &0.5773  &3.7770  &0.5120\\
  7   &2  &5.9856  &0.9944 &10.6969  &1.0000  &8.7545  &0.9997\\
  9   &2  &6.2106  &0.9567  &5.2452  &0.9353  &3.7604  &0.9000\\
 10   &2  &4.1649  &0.2864  &3.7057  &0.3914  &3.3865  &0.3590\\
 11   &2  &5.4050  &0.9999  &9.1705  &1.0000  &5.6743  &0.9993\\
 12   &2  &3.8657  &0.3916  &3.7061  &0.3915  &3.3868  &0.3591\\
 13   &2  &7.3354  &0.9649 &10.6700  &1.0000  &8.7279  &0.9993\\
 14   &2  &6.9360  &0.5401  &9.3649  &1.0000  &6.4017  &0.9973\\
\bf  15   &\bf 2  &0.0284  &0.9992  &\bf 0.0000  &\bf 1.0000  &\bf 0.0000  &\bf 1.0000\\
\hline
\end{tabular}

{\footnotesize\footnotemark[1]Wolfram class.\hfill}
\label{n=12}
\end{table}

\addtocounter{table}{-1}
\begin{table}[p]
\centering
\caption{Continued.}
\begin{tabular}{|c|c|cc|cc|cc|}
\hline
& &\multicolumn{2}{|c|}{$p=0$} &\multicolumn{2}{|c|}{$p=0.001$} &\multicolumn{2}{|c|}{$p=0.01$}\\
\cline{3-8}
Rule &W.\ cl.\footnotemark[1] &$G^\sigma$ &$r^\sigma$ &$G^\pi$ &$r^\pi$ &$G^\pi$ &$r^\pi$\\
\hline
\bf  19   &\bf 2  &\bf 4.4312  &\bf 1.0000 &10.2245  &1.0000  &6.9111  &0.9999\\
\bf  23   &\bf 2  &4.8031  &\underline{1.0000} &\bf 10.6072  &\bf 1.0000  &\bf 8.4314  &\bf 1.0000\\
 24   &2  &6.1840  &0.1898  &5.4814  &0.2167  &4.8971  &0.1797\\
 25   &2  &5.2543  &0.9866  &4.6353  &0.9800  &2.9800  &0.9617\\
 26   &2  &3.2749  &0.6617  &2.3692  &0.8083  &1.7615  &0.7537\\
 27   &2  &2.5162  &0.9884  &2.2044  &0.9847  &1.7538  &0.9824\\
 28   &2  &6.0127  &0.9416 &10.1677  &1.0000  &6.5555  &0.9998\\
\bf  29   &\bf 2  &1.6331  &0.9930  &\bf 1.6753  &\bf 1.0000  &\bf 1.4720  &\bf 1.0000\\
 33   &2  &4.3519  &0.8883  &3.9283  &0.8923  &3.3398  &0.8662\\
 35   &2  &3.9843  &0.9724  &3.9851  &0.8650  &2.9320  &0.8386\\
 36   &2  &8.1266  &0.0737  &6.5768  &0.0638  &6.0508  &0.0455\\
 37   &2  &5.6552  &0.9587  &4.5088  &0.9431  &3.0788  &0.9306\\
 38   &2  &3.4141  &0.4347  &3.0763  &0.4666  &2.6628  &0.4259\\
 41   &2  &5.9904  &0.9579  &4.9622  &0.9321  &2.7975  &0.8301\\
\bf  43   &\bf 2  &5.7458  &0.9999  &\bf 9.4076  &\bf 1.0000  &\bf 6.4673  &\bf 1.0000\\
 46   &2  &6.1879  &0.7170  &5.4573  &0.9625  &4.6255  &0.9581\\
 50   &2  &4.9581  &0.9572 &10.3458  &1.0000  &7.2818  &0.9999\\
\bf  51   &\bf 2  &\bf 0.0000  &\bf 1.0000  &\bf 0.0000  &\bf 1.0000  &\bf 0.0000  &\bf 1.0000\\
\bf  57   &\bf 2  &8.8859  &\underline{1.0000}  &\bf 9.8724  &\bf 1.0000  &\bf 5.7901  &\bf 1.0000\\
 58   &2  &8.8538  &0.8375  &9.1855  &0.9046  &5.7138  &0.9108\\
 62   &2  &5.0852  &0.9783  &4.4037  &0.9323  &2.3477  &0.8830\\
\bf  77   &\bf 2  &4.7242  &0.9682 &\bf 10.6072  &\bf 1.0000  &\bf 8.4314  &\bf 1.0000\\
\hline
\end{tabular}

{\footnotesize\footnotemark[1]Wolfram class.\hfill}
\end{table}

\addtocounter{table}{-1}
\begin{table}[p]
\centering
\caption{Continued.}
\begin{tabular}{|c|c|cc|cc|cc|}
\hline
& &\multicolumn{2}{|c|}{$p=0$} &\multicolumn{2}{|c|}{$p=0.001$} &\multicolumn{2}{|c|}{$p=0.01$}\\
\cline{3-8}
Rule &W.\ cl.\footnotemark[1] &$G^\sigma$ &$r^\sigma$ &$G^\pi$ &$r^\pi$ &$G^\pi$ &$r^\pi$\\
\hline
 94   &2  &4.8790  &0.9885  &8.9317  &0.9058  &4.7377  &0.9148\\
\bf 178   &\bf 2  &4.9578  &0.9573 &\bf 10.6072  &\bf 1.0000  &\bf 8.4314  &\bf 1.0000\\
197   &2  &7.4741  &0.9821 &10.4449  &1.0000  &7.6695  &0.9978\\
\bf 198   &\bf 2  &6.0131  &0.9324 &\bf 10.3312  &\bf 1.0000  &\bf 7.1215  &\bf 1.0000\\
201   &2  &2.6352  &0.5459  &2.7661  &0.4743  &2.3094  &0.4402\\
\bf 204   &\bf 2  &\bf 0.0000  &\bf 1.0000  &\bf 0.0000  &\bf 1.0000  &\bf 0.0000  &\bf 1.0000\\
205   &2  &1.8069  &0.5713  &1.7622  &0.6299  &1.5918  &0.6054\\
210   &2  &0.2001  &0.9861  &0.0620  &0.9996  &0.0363  &0.9996\\
\bf 212   &\bf 2  &6.1928  &0.5834  &\bf 9.4076  &\bf 1.0000  &\bf 6.4673  &\bf 1.0000\\
214   &2  &5.1102  &0.9655  &4.6487  &0.6824  &3.6478  &0.6075\\
217   &2  &5.6595  &0.2863  &5.4256  &0.2331  &4.7625  &0.1972\\
218   &2  &6.7277  &0.9966  &6.0657  &0.1033  &5.5166  &0.0752\\
220   &2  &3.8652  &0.4233  &3.6985  &0.3906  &3.3377  &0.3541\\
222   &2  &4.7980  &0.5614  &4.4711  &0.1503  &4.1845  &0.1271\\
\bf 226   &\bf 2  &4.8855  &0.3928  &\bf 4.8729  &\bf 1.0000  &\bf 3.9663  &\bf 1.0000\\
227   &2  &4.8331  &0.3698  &4.7018  &0.8802  &3.7120  &0.8524\\
228   &2  &5.8775  &0.2538  &5.3899  &0.2298  &4.5433  &0.1823\\
229   &2  &5.0590  &0.3979  &3.9467  &0.6617  &3.1185  &0.5957\\
230   &2  &7.0931  &0.9899  &5.4817  &0.2165  &4.8938  &0.1780\\
\bf 232   &\bf 2  &4.8016  &0.7413 &\bf 10.6072  &\bf 1.0000  &\bf 8.4314  &\bf 1.0000\\
233   &2  &9.4717  &0.2346 &11.8348  &0.0009 &10.7957  &0.0048\\
236   &2  &3.6091  &0.4395 &11.7444  &0.0082 &10.1974  &0.0448\\
\hline
\end{tabular}

{\footnotesize\footnotemark[1]Wolfram class.\hfill}
\end{table}

\addtocounter{table}{-1}
\begin{table}[p]
\centering
\caption{Continued.}
\begin{tabular}{|c|c|cc|cc|cc|}
\hline
& &\multicolumn{2}{|c|}{$p=0$} &\multicolumn{2}{|c|}{$p=0.001$} &\multicolumn{2}{|c|}{$p=0.01$}\\
\cline{3-8}
Rule &W.\ cl.\footnotemark[1] &$G^\sigma$ &$r^\sigma$ &$G^\pi$ &$r^\pi$ &$G^\pi$ &$r^\pi$\\
\hline
237   &2  &8.1243  &0.4170 &11.8230  &0.0015 &10.7351  &0.0077\\
\bf 240   &\bf 2  &0.0387  &0.9186  &\bf 0.0000  &\bf 1.0000  &\bf 0.0000  &\bf 1.0000\\
241   &2  &2.1014  &0.6609  &1.6306  &0.5414  &1.4663  &0.5100\\
242   &2  &4.5260  &0.9234  &3.6120  &0.5058  &3.2760  &0.4794\\
243   &2  &4.1114  &0.2931  &3.7061  &0.3915  &3.3868  &0.3591\\
244   &2  &2.6381  &0.9994  &2.7798  &0.4707  &2.5005  &0.4393\\
246   &2  &6.5741  &0.9102  &5.4668  &0.2351  &5.0369  &0.2019\\
\hline
 18   &3  &5.4850  &0.5728  &4.9312  &0.6297  &3.7099  &0.5390\\
 22   &3  &8.8848  &0.3983  &6.6699  &0.6261  &3.5737  &0.6611\\
 30   &3  &4.6560  &0.8994  &1.7843  &0.9831  &0.3332  &0.9958\\
 45   &3  &7.6936  &0.9999  &1.6755  &1.0000  &0.0998  &0.9999\\
\bf  60   &\bf 3  &4.1853  &0.9940  &\bf 3.5731  &\bf 1.0000  &\bf 1.7232  &\bf 1.0000\\
 73   &3  &4.3257  &0.9635  &4.1166  &0.9965  &2.5287  &0.9893\\
\bf  90   &\bf 3  &4.0586  &0.9980  &\bf 3.7778  &\bf 1.0000  &\bf 2.6528  &\bf 1.0000\\
\bf 105   &\bf 3  &8.1250  &0.9875  &\bf 7.4925  &\bf 1.0000  &\bf 4.7941  &\bf 1.0000\\
126   &3  &5.3877  &0.9935  &4.5984  &0.9971  &3.3678  &0.9891\\
\bf 150   &\bf 3  &8.1250  &0.9895  &\bf 7.4925  &\bf 1.0000  &\bf 4.7941  &\bf 1.0000\\
161   &3  &5.3009  &0.9956  &4.5606  &0.9977  &2.9457  &0.9836\\
182   &3  &5.3737  &0.7992  &5.0251  &0.6066  &3.6983  &0.5182\\
225   &3  &9.6682  &0.2317  &4.4156  &0.1840  &0.5267  &0.9178\\
\hline
 54   &4  &6.5286  &0.8871  &6.6093  &0.9993  &2.8375  &0.9788\\
193   &4  &6.7149  &0.9740  &4.5883  &0.9826  &1.7759  &0.9589\\
\hline
\end{tabular}

{\footnotesize\footnotemark[1]Wolfram class.\hfill}
\end{table}

Except for rule and Wolfram-class identifications, all numbers in
Tables~\ref{n=11} and~\ref{n=12} originate from results that were output by our
programs with six decimal places. Owing to space considerations, in the tables
they are given with four decimal places only. This has caused no rounding
problems in the vast majority of cases, but those cases in which problems did
arise are in the tables highlighted by underlining the corresponding numbers.
There is one occurrence in Table~\ref{n=11} (class-3 rule 30, for which the
table says $r^\pi=1.0000$ for both values of $p$, but these are rounded up from
$r^\pi=0.999988$ and $r^\pi=0.999985$, respectively for $p=0.001$ and $p=0.01$)
and there are two occurrences in Table~\ref{n=12} (class-2 rules 23 and 57, for
which the table says $r^\sigma=1.0000$, but this is rounded up from
$r^\sigma=0.999997$ and $r^\sigma=0.999999$, respectively).

\section{Discussion}\label{disc}

Tables~\ref{n=11} and~\ref{n=12} contain several entries with extremal values of
the quantities they represent, at least as far as can be gleaned from the six
decimal places we printed out. Such extremal values refer in some cases to
situations in which information gain is equal to either $0$ or $n$, as well as
situations in which the total-correlation ratios are equal to $0$, and in most
cases to situations in which the total-correlation ratios are equal to $1$. All
these entries are highlighted in the tables through the use of special
typefaces. Since it turns out that all extremes can be covered by focusing on
minimum ($0$) or maximum ($1$) ratio values, we italicize all gain-ratio pairs
for which the ratio is minimum and use boldface for gain-ratio pairs for which
the ratio is maximum (provided, in the probabilistic case, that the maximum is
observed for both $p=0.001$ and $p=0.01$). The corresponding rule numbers and
class identifications are also modified in this way.

In what follows, we occasionally refer to specific features of a rule's
basin-of-attraction field for a given value of $n$. We do this whenever the
features in question help understand particular values of information gain or
total-correlation ratio. We refer the reader to one of the available atlases
\cite{wl92,wa} for a lookup of such features.

\subsection{Maximum information gain (and minimum total correlation) in the
deterministic case}

By Equation~(\ref{eq:jgain}), we have $G^\sigma=n$ if and only if $H^\sigma=0$.
Achieving $H^\sigma=0$, in turn, is tantamount to the long-run situation in
which $\sigma_i=1$ for some CA state $i$, or equivalently to either
$\sigma_{c,0}=1$ or $\sigma_{c,1}=1$ for each cell $c$. The latter happens if
and only if $H^\sigma_c=0$ for every cell $c$, which by
Equation~(\ref{eq:tcorr}) implies $C^\sigma=0$ and, of course, $r^\sigma=0$ as
well.

By Equation~(\ref{eq:det}), the condition that $\sigma_i=1$ for some CA state
$i$ holds if and only if the CA being considered entails one single basin of
attraction, encompassing all $2^n$ states, and moreover this basin's attractor
is a fixed point (one single state to which the CA dynamics recurs perpetually
once it is reached). In fact, this is what we see in Table~\ref{n=11} (for
class-1 rules 0, 251, and 253) and in Table~\ref{n=12} (for rules 0 and 253
only, as for $n=12$ rule 251 leads to the appearance of a further basin).

It is curious to note that, for deterministic scenarios in which such
single-basin, single-state-attractor condition holds, the resulting
$H^\sigma_c=0$ for every cell $c$ implies $H^\sigma=0$ as well, since as noted
in Section~\ref{ii}, we always have $\sum_{c=1}^nH^\sigma_c\ge H^\sigma$. So,
when $r^\sigma=0$ (or, equivalently, $C^\sigma=0$ with $G^\sigma>0$), and thus
$\sum_{c=1}^nH^\sigma_c=H^\sigma$, any of these deterministic scenarios implies,
through $H^\sigma=0$, that $G^\sigma=n$. In summary, not only do the rules
singled out above imply $r^\sigma=0$, they are the only ones to do so for the
two values of $n$ being discussed.

\subsection{Minimum information gain in the deterministic case}

Resorting once again to Equation~(\ref{eq:jgain}), we see that having
$G^\sigma=0$ is equivalent to having $H^\sigma=n$, that is, a long-run scenario
of maximum possible uncertainty. Naturally, $H^\sigma=n$ happens if and only if
$\sigma_i=1/2^n$ for every CA state $i$, which by Equation~(\ref{eq:det}) is
equivalent to all CA states being periodic. This is the case precisely for
class-2 rules 51 and 204 (which, not coincidentally, are complementary to each
other), as seen in both Table~\ref{n=11} and Table~\ref{n=12}.

It also follows easily from Equation~(\ref{eq:fgain}) that $G^\sigma=0$ implies
$C^\sigma=0$, whence $r^\sigma=1$ (by convention, since total correlation is in
this case $100\%$ of all the information gain), which is reflected in the two
tables as well. As we note below, however, depending on the rule under
consideration and on the value of $n$, it is entirely possible to have
$r^\sigma=1$ also for $G^\sigma>0$.

\subsection{Minimum information gain in the probabilistic case}

We observed no occurrence of $G^\pi=n$, which as far as we can tell suggests
that, once probabilistic deviations from the deterministic CA recipe are
allowed, the CA dynamics works to prevent the uncertainty regarding the long-run
CA state from being too low. Contrasting with this, the opposite extreme of
$G^\pi=0$ does occur sometimes, which by now we easily associate with $H^\pi=n$
and a uniform distribution $\pi$ (i.e., $\pi_i=1/2^n$ for every CA state $i$).
As is well known, this happens if and only if the transition-probability matrix
$P$ is doubly stochastic: not only do its rows add up to $1$
[cf.\ Equation~(\ref{eq:stoch})], but so do its columns.

Because the elements of $P$, the $p_{i,j}$'s, are given as in
Equation~(\ref{eq:tprob1}), one simple shortcut toward double stochasticity is
to let $p=0.5$, yielding $p_{i,j}=1/2^n$ for all $i,j$. Another, unrelated way
is to let $p_{i,j}=1/{n\choose\tau}$ if $D_{i,j}=\tau$ (if not, then
$p_{i,j}=0$), where $\tau\le n$ is any number of cells \cite{vb04}. These two
examples lead to a symmetric $P$, i.e., to $p_{i,j}=p_{j,i}$ for all $i,j$.
Clearly, whenever this happens, $P$ is seen to be doubly stochastic simply by
virtue of being stochastic in the first place. We note, however, that it is
possible for a stochastic matrix to be doubly stochastic without being
symmetric.

In Table~\ref{n=11}, for $n=11$, we find $G^\pi=0$ for class-2 rules 15, 51,
204, 210, and 240, and for class-3 rules 45, 105, and 150, all of which can be
grouped into complementary pairs: 15 with 240, 51 with 204, 45 with 210, 105
with 150. All corresponding matrices are indeed doubly stochastic, and in
particular those of rules 51 and 204 are symmetric. As for $n=12$, four of these
same rules are also those for which $G^\pi=0$ in Table~\ref{n=12}, namely rules
15, 51, 204, and 240. All four result in doubly-stochastic matrices, and again
the matrices for rules 51 and 204 are symmetric.

As in the case of $G^\sigma=0$ above, we recognize that $G^\pi=0$ implies
$C^\pi=0$, and therefore $r^\pi=1$ (once again by convention, and once again we
note that $r^\pi=1$ also happens in situations of $G^\pi>0$, as we discuss
below).

\subsection{Maximum total correlation in the deterministic case}

We get $r^\sigma=1$ in Equation~(\ref{eq:fratio}) if and only if, by
Equation~(\ref{eq:fgain}), $\sum_{c=1}^nG^\sigma_c=0$. This, in turn, is by
Equation~(\ref{eq:mgain}) equivalent to having $H^\sigma_c=1$ for every cell
$c$, or to having cell-state probabilities $\sigma_{c,0}=\sigma_{c,1}=0.5$
regardless of $c$. By Equation~(\ref{eq:dmprob}), this happens if in every basin
the attractor has an even number of CA states along which every cell has state
$0$ as often as it has state $1$, but conceivably this may not be necessary
(i.e., depending on the rule and on the value of $n$, other basin and attractor
arrangements may exist that lead to the desired probabilities).

If we ignore the already noted rules 51 and 204 (which lead to $r^\sigma=1$ only
degenerately, by virtue of doing so for $G^\sigma=0$), Tables~\ref{n=11}
and~\ref{n=12} provide us with further rules, as follows. For $n=11$, class-2
rules 11, 15, 19, and 23, as well as class-3 rule 105. For $n=12$, rule 19 only.
Of these, only rule 11 does not conform to the simple sufficient condition we
outlined, so either a subtler arrangement is at play or we really have
$r^\sigma<1$ but missed this fact because of insufficient precision in the
numbers that were output. In this case we favor the latter hypothesis, since for
$n=11$ this rule's basin-of-attraction field has only one basin (out of $11$)
for which the number of periodic states is odd, which may for example disrupt
the sufficient condition enough to prevent $r^\sigma$ from being exactly $1$.

\subsection{Maximum total correlation in the probabilistic case}

Similarly to the deterministic case discussed above, the necessary and
sufficient condition for obtaining $r^\pi=1$ is that
$\sigma_{c,0}=\sigma_{c,1}=0.5$ for every cell $c$. That is, in the long run
every cell is as likely to be found in state $0$ as it is to be found in state
$1$.

Tables~\ref{n=11} and~\ref{n=12} reveal several rules for which $r^\pi=1$, even
if we ignore all those that, as noted earlier, have $r^\pi=1$ only as a
consequence of $G^\pi=0$. For $n=11$, the further rules are class-2 rules 23,
29, 43, 57, 77, 178, 198, 212, 226, and 232, and also the class-3 rules 45, 60,
and 90 (the XOR rule). These are also the further rules for $n=12$, but now
joined by class-3 rules 105 and 150. These rules' basin-of-attraction fields are
richly assorted and no pattern seems to emerge that might help explain why they
promote maximum total correlation for the values of $n$ and $p$ in use.

\subsection{Minimum total correlation in the probabilistic case}

Unlike the case of minimum total correlation in deterministic scenarios
discussed above, in which $r^\sigma=0$ implies $G^\sigma=n$, here there is no
reason to expect that $r^\pi=0$ (or, equivalently, $C^\pi=0$ with $G^\pi>0$)
should imply $G^\pi=n$. In fact, in Tables~\ref{n=11} and~\ref{n=12} we see that
$G^\pi<n$ for all rules displaying $r^\pi=0$. Interestingly, all (and only)
class-1 rules are such that $r^\pi=0$ in the tables.

The basin-of-attraction fields of all class-1 rules are characterized by the
concentration of nearly all CA states in a single basin (all of them, in the
cases noted earlier), this one basin having a single-state attractor at its
core. As we solve for the various $\pi_i$'s with $p\ll 0.5$, this attractor
state gets most of the probability mass while that of the others is very small.
As a result, in the long run the CA is found in the situation of having a
relatively small entropy $H^\pi$, and marginal entropies $H^\pi_c$ whose sum
over all cells is above $H^\pi$ by only a negligible margin. Thus, a value for
$C^\pi$ is obtained that is indistinguishable from $0$ (at least within the
precision we adopted). Naturally, the small value of $H^\pi$ is precisely the
amount by which $G^\pi$ falls short of equaling $n$.

\subsection{Total correlation tends to be unexpectedly high}

Tables~\ref{n=11} and~\ref{n=12} do not contain explicit values of total
correlation, but these can be easily estimated by resorting to the simple
relation given by Equation~(\ref{eq:fratio}), $C^\rho=r^\rho G^\rho$. If we
ignore those (arguably degenerate) cases of zero gain (and thus zero total
correlation), a fairly simple inspection reveals that only for a few rules do we
have $C^\sigma<0.6$ or $C^\pi<0.6$ (for either value of $p$) for both $n=11$ and
$n=12$. These are class-2 rules 4, 36, 218, 222, 233, 236, and 237, along with
class-3 rules 30 and 225.

The choice of the $0.6$ threshold, though somewhat arbitrary as will become
clear shortly, ultimately has to do with how much uncertainty would be expected,
in the long run, if every possible probability distribution over the $2^n$ CA
states were taken into account. In other words, the question is, what is the
expected value of $H^\rho$ over all possible probability distributions $\rho$?
The answer to this question depends on what weight each of these infinitely many
probability distributions is to have when computing the desired expected value.
If we assume that all weights are to be the same (that is, the probability
density to be used over all distributions is uniform), then it is a known fact
that the expected value of $H^\rho$ can be well approximated by
$n-(1-\gamma)/\ln 2$ \cite{c95}, where $\gamma\approx 0.57722$ is the
well-known Euler constant, even for values of $n$ as modest as the ones we have
been using. By Equation~(\ref{eq:jgain}), the expected value of the information
gain $G^\rho$ can itself be approximated by $(1-\gamma)/\ln 2\approx 0.6$, which
by Equation~(\ref{eq:fgain}) can be taken as an upper bound on the expected
value of total correlation $C^\rho$.

So, by pinpointing those rules for which $C^\rho$ falls below this upper bound
of about $0.6$ for at least one of the $\rho$'s of interest ($\sigma$ or one of
the $\pi$'s), we are singling out rules for which $C^\rho$ may lie above the
actual expected value just as it may lie below it. Even so, the resulting number
of rules is surprisingly low (only nine all told). Taking this together with the
nondegenerate cases of $r^\pi=1$ discussed above, elementary CA seem to perform
remarkably well in generating information gain beyond that which is generated
individually by the cells.

In this respect, we find it instructive to single-out class-2 rule 7, for which
total correlation is highest in either table, specifically $C^\pi=9.7352$ for
$n=11$ and $C^\pi=10.6969$ for $n=12$, with $p=0.001$ in both cases. These two
figures match the entirety of the corresponding information gains as far as we
can tell from the available decimal places, so $r^\pi=1$, though rule 7 is not
highlighted in either table because our highlighting criterion requires maximum
ratios for the two values of $p$. A similar case is that of class-2 rule 13 for
$n=12$ and $p=0.001$. This rule has $C^\pi=10.67$ and ranks second in
Table~\ref{n=12} for total correlation, also with $r^\pi=1$.

\subsection{Deterministic versus probabilistic scenarios}

The scenario we have been calling deterministic is the traditional CA scenario
in which CA state $i$ is followed by CA state $k_i$ at the next time step with
probability $1$. Despite its denomination, the deterministic scenario is subject
to uncertainty (quantified, e.g., through the information gain $G^\sigma$)
regarding the long-run state in which the CA is to be found, since the initial
CA state is chosen randomly.

As we noted in Section~\ref{model}, the deterministic scenario can be thought of
as the special case of the probabilistic scenario that sets $p$ to $0$, as by
Equation~(\ref{eq:tprob1}) in this case we get $p_{i,j}=1$ if $j=k_i$. We also
noted, earlier in Section~\ref{disc}, that setting $p$ to $0.5$ is the same as
obtaining the minimum possible gain in the probabilistic scenario (i.e.,
$G^\pi=0$). We might then be led to believe that, regarding the evolution of
information gain as $p$ is varied, changing $p$ through an increasing sequence
from $p=0$ toward $p=0.5$ would reveal a succession of ever-decreasing gain
values: first $G^\sigma$ (for $p=0$); then a succession of $G^\pi$ values (for
strictly positive values of $p$), each one surpassing neither its predecessor
nor $G^\sigma$.

This is indeed what we often find as we scan the rows of Tables~\ref{n=11}
and~\ref{n=12} from left to right, but not always. The exceptions are not too
numerous, but one of them is particularly striking because
$G^\pi/G^\sigma\approx 3.254$ for $p=0.001$, meaning that for this rule and the
two values of $n$ under consideration, information gain more than triples as we
move from the deterministic case to the probabilistic one with $p=0.001$. The
rule in question, in either table, is class-2 rule 236. It is reassuring,
however, that nowhere do we find an increase in $G^\pi$ as $p$ is increased,
because this we can expect in a principled manner: increasing $p$ lets the CA
dynamics deviate from the traditional one ever more and thus produces more
long-run uncertainty (less gain).

Expecting $G^\sigma>G^\pi$ to always hold is unreasonable, though, because the
deterministic scenario is a special case of the probabilistic one only insofar
as the transition probabilities $p_{i,j}$ are concerned. For $p>0$, every CA
state is reachable from every other (and from itself) in one step during the CA
dynamics, albeit in most cases with low probability. For $p=0$, on the other
hand, only $k_i$ can be reached from CA state $i$ in one time step. As we noted
in Section~\ref{model}, this affects the method used to find $\sigma$ or $\pi$
profoundly [simply applying Equation~(\ref{eq:det}), in the former case, and
finding a Markov chain's stationary probabilities, in the latter]. More
tellingly, it abruptly affects the structural possibilities for the long-run
mix-up of CA states at the boundary between $p=0$ and $p>0$. As a consequence,
comparing $G^\sigma$ and $G^\pi$ seems insufficiently principled in general.

\subsection{Class-4 rules and total correlation}

With the exception of class-1 rules, which hardly display any total correlation
for the various reasons we have noted, all highlights in Tables~\ref{n=11}
and~\ref{n=12} refer to the achievement of maximum total correlation, that is,
total correlation that accounts for all the information gain. The rules in
question are all class-2 or class-3 rules, which characteristically behave in
such a way as to produce spatiotemporal patterns often referred to as ``dull''
and ``chaotic,'' respectively. But what of the two class-4 rules, namely rule
54 and rule 193 (this one equivalent by both negation and reflection to the
famous rule 110, provably capable of universal computation)? Can they not
generate substantial total correlation as well?

The answer is that they can, but without coming near some of the top-ranking
rules we have encountered. Although their total-correlation ratios are above
$0.9$ almost always (the exceptions being $r^\sigma$ for rule 193 with $n=11$
and much less severely for rule 54 with $n=12$), the information gains are less
than impressive and thus so are the total correlations themselves. So, at least
in the case of rule 193 and perhaps not surprisingly, it appears that being able
to perform universal computation is much more than the ability to integrate
information, perhaps even more stringently than total correlation is much more
than the total information gain the cells are capable of generating locally.

A useful, somewhat quantitative insight into the issue of relatively modest
information gains can be had through one of the chiefest characteristics
normally attributed to class-4 rules, namely that their basin-of-attraction
fields are dominated by long transients leading to not too short, or long,
attractors. In the deterministic case this suggests that the distribution
$\sigma$ is probably neither too concentrated on very few states nor spread out
to the point of resembling the uniform distribution. Such a $\sigma$, as we
know, leads to mid-valued $H^\sigma$ and $G^\sigma$. In the probabilistic case,
too, such a purportedly typical layout of a class-4 field can be influential. By
Equation~(\ref{eq:tprob2}), and so long as $p<0.5$, CA state $i$ is always far
more likely to be succeeded by state $k_i$ than by any other. This suggests, for
distribution $\pi$, properties similar to those of $\sigma$, with similar
effects on $H^\pi$ and thence on $G^\pi$.

\section{Conclusion}\label{concl}

We have studied information integration in elementary CA, using the notion of
total correlation and its relation to information gain as guiding principles.
These entities are mathematical functions of a system's random elements, which
in the case of CA requires that they be extended by some sort of nondeterminism.
We have done this in two different ways, one in which a CA's basic determinism
is preserved but its initial state is chosen uniformly at random, another in
which every cell is prone to disobeying the deterministic rule that governs its
behavior probabilistically and independently of all others. Both sources of
nondeterminism lead to long-run probability distributions on the CA states and
those can be used to compute total correlation.

What is tantalizing about total correlation is that it hints at the existence
of ways to reduce uncertainty that emerge out of the interaction of the
underlying system's components. That is, even though such components can create
information through their evolution in time, often there is additional
information that is created by how the components interact with one another. In
the case of elementary CA we have identified rules in Wolfram classes 2 and 3
that excel at creating such additional information. For these rules, most (if
not all) of the information that is created is of the total-correlation type,
turning the rules themselves into possible models of information integration.

Our results are preliminary in several regards, particularly in regard to the
fact that they refer to the simplest possible CA and in regard to the fact that,
by virtue of the exponential growths that typically characterize CA studies,
only for small systems have we been able to obtain numerical results. Especially
useful headways can be expected from provably correct approximations to the
Markov-chain solution methods if some simplifying structure in the transition
matrix comes to be identified, and also from making theoretical progress toward
understanding how total correlation behaves for certain classes of CA rules. To
the best of our knowledge, however, such goals are still elusive at this time.

We close by commenting on a very apt note, by J.\ Rothstein in 1952, regarding
the notion of organization \cite{r52}. In this note, what the author calls an
``alternative'' can be identified with a random variable (a cell's state in
our case). A ``selection'' is a value assignment to this random variable, and
likewise a ``complexion'' is a joint value assignment to all variables (a ``set
of selections''). The heart of the note, as we see it, is the author's
observation that, in the general case, the ``entropy of the set of complexions
is $\ldots$ less than the sum of the entropies of the sets of alternatives.''
This, in our view, came as close to foreshadowing the concept of total
correlation as can be imagined. Of course, in the meantime since then it took
the development of information theory and of computing technology for some of
the concept's potential to be realized and some of its consequences to be
understood. Sometime in the future, likewise, the purported link between total
correlation and elusive entities like consciousness may come to be clarified. We
believe the present work can contribute to such developments by having
demonstrated, though to a limited extent, that CA can be used as simple
computational models of how information can be efficiently integrated.

\subsection*{Acknowledgments}

We acknowledge partial support from CNPq, CAPES, a FAPERJ BBP grant, and the
joint PRONEX initiative of CNPq and FAPERJ under contract E-26/110.550/2010.

\bibliography{iica}
\bibliographystyle{plain}

\end{document}